\documentclass[prl,twocolumn,floatfix,superscriptaddress]{revtex4-1}
\usepackage{dcolumn,amsmath}
\usepackage{graphicx}
\usepackage{bm}
\usepackage{hyperref}

\newcommand{\ecm}{\ensuremath{e {\cdotp} {\rm cm}}}
\newcommand{\eEDM}{{\em e}EDM}
 \newcommand{\Eeff}{\mathcal{E}_\mathrm{eff}}
\newcommand{\Bvec}{\vec{\mathcal{B}}} 
\newcommand{\Evec}{\vec{\mathcal{E}}} 

\newcommand{\de}{d_\mathrm{e}}

\begin{document}
   \title{The Rabi frequency on the $H^3\Delta_1$ to $C^1\Pi$ transition in ThO: influence of interaction with electric and magnetic fields}
\author{A.N.\ Petrov}\email{alexsandernp@gmail.com}
\affiliation
{National Research Centre ``Kurchatov Institute'' B.P. Konstantinov Petersburg
Nuclear Physics Institute, Gatchina, Leningrad district 188300, Russia}
\affiliation{Department of Physics, St. Petersburg University,7/9 Universitetskaya nab., St. Petersburg, 199034 Russia}
\begin{abstract}
Calculations of the correlations between the Rabi frequency on the $H^3\Delta_1$ to $C^1\Pi$ transition in ThO molecule and experimental setup parameters
in the electron electric dipole moment (\eEDM) search experiment is performed. Calculations are required for estimations of systematic errors in the experiment
due to imperfections in laser beams used to prepare the molecule and read out the \eEDM\ signal.%
\end{abstract}

\maketitle

The current limit for the electron electric dipole moment (\eEDM), $|\de|<9\times 10^{-29}$ \ecm\ (90\% confidence), was set by measuring the spin precession  
of thorium monoxide (ThO) molecules in the metastable electronic $H^3\Delta_1$ state \cite{Baron2013}.
The measurements were performed on the ground rotational level which has two closely-spaced $\Omega$-doublet levels of opposite parity.
It was shown that due to existence of closely-spaced $\Omega$-doublet levels the experiment on ThO is very robust against a number of systematic effects \cite{DeMille2001, Petrov:14, Vutha2009, Petrov:15}.
Both the state preparation and the spin precession angle, $\phi$, measurement is performed by optically pumping the  $H^3\Delta_1 \rightarrow C^1\Pi$
transition with linearly polarized laser beam \cite{Baron2013}. The transition to ground rotational level of $C^1\Pi$ which has similar to $H^3\Delta_1$ $\Omega$-doublet structure (see below)
is used. 
Searching for systematic errors is an important part of the \eEDM~ search experiment. It was found that the dominant systematic errors in the experiment \cite{Baron2013} are due to imperfections in laser beams used to
prepare the molecule and read out the \eEDM~ signal \cite{ACME:17}. In particular, it was found that the spin precession angle $\phi$ has additional systematic contribution $\Phi$ due to small
changes of the Rabi frequency  $\frac{d\Omega_r}{\Omega_r}$:
\begin{eqnarray}
\Phi = \beta \frac{d\Omega_r}{\Omega_r},
\label{system}
\end{eqnarray}
where $\beta \sim 10^{-3}$ \cite{ACME:17}.
The measurement of spin precession is repeated under different conditions which can be characterized by binary parameters being switched from $+1$ to $-1$. 
The three primary binary parameters are $\tilde {\cal N}$, $\tilde {\cal E}$, $\tilde {\cal B}$.
${\cal \tilde{N}} {=} +1(-1)$ means that the measurement was performed for lower (upper) $\Omega$-doublet level of $H^3\Delta_1$.
$\tilde {\cal E} = {\rm sgn}(\hat{z}\cdot\Evec)$ and $\tilde {\cal B} = {\rm sgn}(\hat{z}\cdot\Bvec)$ define the orientation of the external static electric and magnetic
fields respectively along the laboratory axis $z$.
The measured precession angle $\phi$ can be represented as \cite{ACME:17}

\begin{eqnarray}
\nonumber
\phi({\cal \tilde{N}},{\cal \tilde{E}},{\cal \tilde{B}}) = \phi^{nr} 
+{{\cal \tilde{B}}}\phi^{{\cal {B}}}
+{{\cal \tilde{E}}}\phi^{{\cal {E}}}
+{{\cal \tilde{E}}{\cal \tilde{B}}}\phi^{{\cal {E}}{\cal {B}}} \\
\label{phiNEB}
+{{\cal \tilde{N}}}\phi^{{\cal {N}}}
+{{\cal \tilde{N}}{\cal \tilde{B}}}\phi^{{\cal {N}}{\cal {B}}}
+{{\cal \tilde{N}}{\cal \tilde{E}}}\phi^{{\cal {N}}{\cal {E}}}
+{{\cal \tilde{N}}{\cal \tilde{E}}{\cal \tilde{B}}}\phi^{{\cal {N}}{\cal {E}}{\cal {B}}},
\end{eqnarray}
where notation $\phi^{S_1,S_2...}$ denotes a component which is odd under the switches $S_1,S_2,...$; $\phi^{nr}$
is a component which is even (unchanged) under any of the switches. The \eEDM\ signal is extracted from the ${{\cal \tilde{N}}{\cal \tilde{E}}}$-correlated 
component of the measured phase, $\phi^{{\cal {N}}{\cal {E}}}=\de\Eeff \tau$ \cite{Baron2013}, where
$\Eeff=78~ {\rm GV/cm}$ \cite{Skripnikov:13c,Skripnikov:15a,Skripnikov:16b} is the effective electric field acting on \eEDM\ in the molecule, $\tau$ is interaction time.
In case of an ideal experiment only the $\Omega_r^{nr}$ component is nonzero. However, changes in the preparation and readout laser power
correlated with the switch parameters can lead to nonzero $\Omega_r^{S_1,S_2...}$ components.
In particular, the ${{\cal \tilde{N}}{\cal \tilde{E}}}$-correlated component, according to eq. (\ref{system}), gives rise to systematic errors in the \eEDM\ measurement.
The aim of the present work is to consider $\Omega_r^{S_1,S_2...}$ components which arise due to various perturbations in the $H^3\Delta_1$ and $C^1\Pi$ states.

The basis set describing the $H^3\Delta_1$ and $C^1\Pi$ states wave functions can be presented as product of electronic and rotational wavefunctions $\Psi_{H(C)\Omega}\theta^{J}_{M,\Omega}(\alpha,\beta)$. Here $\Psi_{H(C)\Omega}$ is the electronic wavefunction of the $H^3\Delta_1$ ($C^1\Pi$) state, $\theta^{J}_{M,\Omega}(\alpha,\beta)=\sqrt{(2J+1)/{4\pi}}D^{J}_{M,\Omega}(\alpha,\beta,\gamma=0)$ is the rotational wavefunction, $\alpha,\beta,\gamma$ are Euler angles, and $M$ $(\Omega = \pm 1)$ is the projection of the molecular angular momentum ${\bf J}$ on the laboratory $\hat{z}$ (internuclear $\hat{n}$) axis. For short, we will designate the basis set as $\left|H(C),J, M,\Omega\right>$.
In this paper the $\left|H,J=1,\Omega, M = \pm 1\right>$ and $\left|C,J=1,\Omega, M = 0\right>$ states which are of interest for \eEDM\ search experiment are considered. 

In the absence of external electric field each rotational level splits into two sublevels,
called $\Omega$-doublet levels.
One of them is even $({\cal \tilde{P}}=1)$ and the another one is odd $({\cal \tilde{P}}=-1)$ with respect
to changing the sign of electronic and nuclear coordinates.
The states with ${\cal \tilde{P}}=(-1)^J$  
denoted as $e$ and with ${\cal \tilde{P}}=(-1)^{J+1}$ denoted as $f$ are the 
linear combination of the states with opposite sign of $\Omega$:

\begin{eqnarray}
\nonumber
\left|H(C)J,{\cal \tilde{P}},M\right> = \\
\frac{\left|H(C),J,1,M\right> \pm (-1)^J {\cal \tilde{P}} \left|H(C),J,-1,M\right>} {\sqrt{2}}.
\end{eqnarray}
The experimental values of the $\Omega$-doubling, $\Delta(J)= E(\left|e,J,M\right>) - E(\left|f,J,M\right>)$ are
$\Delta_H=+0.181\,J(J+1)$~MHz for $\left|H\right>$ and $\Delta_C = -25\,J(J+1)$~MHz for $\left|C\right>$ states correspondingly \cite{Baron2013}.

External electric field $\Evec = {\cal \tilde{E}}{\cal E}\hat{z}$ does not couple the
$\left|C,J{=}1,{\cal \tilde{P}}=-1,M{=}0\right>$ and $\left|C,J{=}1,{\cal \tilde{P}}=+1,M{=}0\right>$ states, whereas the $\left|H,J{=}1,{\cal \tilde{P}}=-1,M{=}\pm1\right>$ and $\left|H,J{=}1,{\cal \tilde{P}}=+1,M{=}\pm1\right>$ are coupled:

\begin{eqnarray}
\nonumber
\left|H,{\cal \tilde{E}}, {\cal \tilde{N}}, M\right> = k(-{\tilde {\cal N}})\left|H,J{=}1,{\cal \tilde{P}}{=}-1,M{=}\pm1\right> \\
 - k(+{\tilde {\cal N}}){\cal \tilde{E}}{\cal \tilde{N}}M\left|H,J{=}1,{\cal \tilde{P}}{=}+1,M{=}\pm1\right> ,
\end{eqnarray}
where

\begin{equation}
k(\pm1) = \frac{1}{\sqrt{2}}\sqrt{1 \pm \frac{\Delta_H(J{=}1)}{\sqrt{\Delta_H(J{=}1)^2 + (d_H {\cal E})^2}} },
\label{RK}
\end{equation}
$d_H = -1.612$ a.u. is the dipole moment for $H$ state \cite{Vutha2011, Hess2014Thesis},
${\cal E}>0$ is the magnitude of electric field, ${\cal \tilde{E}}$ defines direction of electric field.

\begin{figure}
\includegraphics[width=3.3 in]{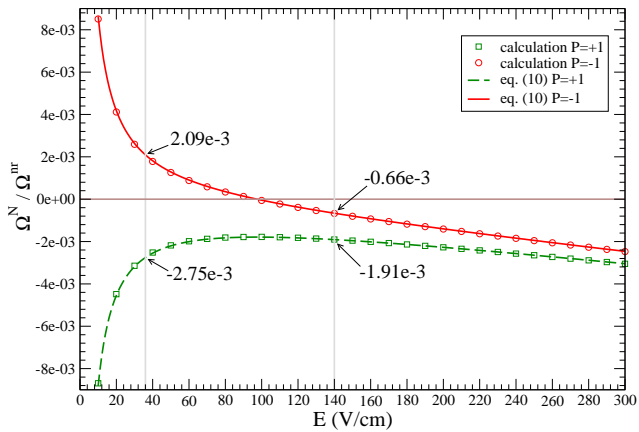}
\caption{(Color online) The Rabi frequency magnitude of the $H$ to $C$ transition. Vertical lines correspond to electric fields ${\cal E} = 38 {\rm ~and~} 140~ {\rm V/cm}$ 
used in the experiment.}
\label{eqnum}
\end{figure}

Then (disregarding the presently unimportant constant) the Rabi frequency on the $H$ to $C$ transition for linearly polarized along the $x$ axis laser beam is

\begin{eqnarray}
\nonumber
\Omega_r^0({\cal \tilde{P}}=+1) = \\
\nonumber
\left<H,{\cal \tilde{E}}, {\cal \tilde{N}}, M\right|x\left|C,J{=}1,{\cal \tilde{P}}=+1,M{=}0\right> \\ 
\label{Om0+1}
 = \frac{\sqrt{2}}{4}d_{HC}k(-{\tilde {\cal N}}),
\end{eqnarray}

\begin{eqnarray}
\nonumber
\Omega_r^0({\cal \tilde{P}}=-1) = \\
\nonumber
\left<H,{\cal \tilde{E}}, {\cal \tilde{N}}, M\right|x\left|C,J{=}1,{\cal \tilde{P}}=-1,M{=}0\right> \\ 
\label{Om0-1}
 = -\frac{\sqrt{2}}{4}d_{HC}{\tilde{\cal N}}{\tilde{\cal E}}k(+{\tilde {\cal N}}),
\end{eqnarray}
where $d_{HC}$ is $H$ to $C$ transition dipole moment. Eqs. (\ref{Om0+1},\ref{Om0-1}) do not take into account interaction with other electronic and rotational states.
Using the angular momentum algebra \cite{LL77}, one can calculate that accounting for Stark mixing between $J{=}1$ and $J{=}2$ rotational levels in $H$ and $C$ states within the first order perturbation theory
gives additional contribution to the Rabi frequency,

\begin{eqnarray}
\nonumber
\Omega_r({\cal \tilde{P}}=+1) = \Omega_r^0({\cal \tilde{P}}=+1) \\
 - {\tilde {\cal N}}k(+{\tilde {\cal N}})\left(\frac{\sqrt{2}}{80} \frac{d_{HC}d_C{\cal {E}}}{B_C} - \frac{3}{80\sqrt{2}} \frac{d_{HC}d_H{\cal {E}}}{B_H}\right),
\label{Om+1}
\end{eqnarray}

\begin{eqnarray}
\nonumber
\Omega_r({\cal \tilde{P}}=-1) = \Omega_r^0({\cal \tilde{P}}=-1) \\
 + {\tilde {\cal E}}k(-{\tilde {\cal N}})\left(\frac{\sqrt{2}}{80} \frac{d_{HC}d_C{\cal {E}}}{B_C} - \frac{3}{80\sqrt{2}} \frac{d_{HC}d_H{\cal {E}}}{B_H}\right),
\label{Om-1}
\end{eqnarray}
where $B_H = 0.32638~ \rm{cm}^{-1}$ and $B_C = 0.322~ \rm{cm}^{-1}$ are rotational constants for $H$ and $C$ states \cite{Edvinsson:84}, $d_C = -1$ a.u. is dipole moment for $C$ state \cite{Hess2014Thesis}.
Eqs. (\ref{Om+1},\ref{Om-1}) give nonzero ${\tilde {\cal N}}$-correlated component of the Rabi frequency

\begin{equation}
\frac{\Omega_r^{\cal N}({\cal \tilde{P}})}{\Omega_r^{nr}} = 
\frac { R(+1, \tilde{\cal P}) - R(-1, \tilde{\cal P})}   { R(+1, \tilde{\cal P}) + R(-1, \tilde{\cal P}) },
\label{Ncor}
\end{equation}

\begin{equation}
R(\tilde {\cal N}, \tilde{\cal P}) = k(-{\tilde {\cal N}}{\tilde {\cal P}}) - 
{\tilde {\cal N}}k(+{\tilde {\cal N}}{\tilde {\cal P}}) 
\left(  \frac{2}{40} \frac{d_C{\cal {E}}}{B_C} - \frac{3}{40} \frac{d_H{\cal {E}}}{B_H}    \right).
\end{equation}
Accounting for interaction with other electronic and rotational states, magnetic field, non ideal laser polarization can further modify eqs. (\ref{Om+1},\ref{Om-1}) and give rise
to correlation of the  Rabi frequency to other switch parameters. To calculate possible correlations for the  Rabi frequency
the numerical calculation was performed.
Following the computational scheme of \cite{Petrov:11, Petrov:14, Petrov:15}, wavefunctions of $H$ and $C$ states
in external {\it static} electric and magnetic fields are obtained by numerical diagonalization of the molecular Hamiltonian over the basis set of the electronic-rotational wavefunctions. Detailed features of the Hamiltonian are described in \cite{Petrov:14}. Comparison of numerical calculations and eq. (\ref{Ncor}) is given in Fig. (\ref{eqnum}). Calculations show that
accounting for perturbations described above does not lead to notable changes in $\frac{\Omega_r^{\cal N}({\cal \tilde{P}})}{\Omega_r^{nr}}$.
One sees that $\frac{\Omega_r^{\cal N}({\cal \tilde{P}})}{\Omega_r^{nr}}$ calculated at the electric fields ${\cal E} = 38 {\rm ~and~} 140~ {\rm V/cm}$ 
used in the experiment are comparable to  $\frac{\Omega_r^{\cal N}({\cal \tilde{P}})}{\Omega_r^{nr}} \approx 2.5 \cdot 10^{-3}$ due to detected laser power correlation  \cite{ACME:17}.
Though, as stated above, external electric field does not couple the
$\left|C,J{=}1,{\cal \tilde{P}}=-1,M{=}0\right>$ and $\left|C,J{=}1,{\cal \tilde{P}}=+1,M{=}0\right>$ states, in the presence of both electric and magnetic
fields the states are coupled. The latter,  together with non ideal laser polarization, leads to ${{\cal \tilde{N}}{\cal \tilde{B}}}$ correlation of the Rabi frequency. 
Zeeman coupling of $J{=}1$ and $J{=}2$ levels of $C$ state leads to ${{\cal \tilde{B}}}$ correlation.
Calculations show that 

\begin{eqnarray}
\label{NBcor}
\frac{\Omega_r^{\cal NB}({\cal \tilde{P}})}{\Omega_r^{nr}} = \frac{-7.4\cdot10^{-7}}{G(V/cm){\rm rad}}{\cal \tilde{P}}{\cal E}{\cal B}d\Theta, \\ 
\label{Bcor}
\frac{\Omega_r^{\cal B}({\cal \tilde{P}})}{\Omega_r^{nr}} = \frac{1.4\cdot10^{-5}}{G\cdot{\rm rad}}{\cal {\cal B}}d\Theta,
\end{eqnarray}
where $d\Theta = \Theta-\pi/4$, $\Theta$ is the elipticity angle which defines laser polarization 
$\hat{\epsilon} =\frac{ \cos{\Theta}(\hat{x} + i\hat{y}) + \sin{\Theta}(\hat{x} - i\hat{y}) } {\sqrt{2}}$.
Deviation of the laser pointing vector $\hat{k}$ from the $\hat{z}$ direction does not modify eqs. (\ref{NBcor},\ref{Bcor}).
$\frac{\Omega_r^{\cal NB}({\cal \tilde{P}})}{\Omega_r^{nr}}$
is suppressed by the relatively large $\Omega$-doubling of the $C$ state.
Other correlations are several orders of magnitude less than $\frac{\Omega_r^{\cal NB}({\cal \tilde{P}})}{\Omega_r^{nr}}$ and $\frac{\Omega_r^{\cal B}({\cal \tilde{P}})}{\Omega_r^{nr}}$
for fields used in the experiment. In particular, $\frac{\Omega_r^{\cal NE}({\cal \tilde{P}})}{\Omega_r^{nr}} < 10^{-14}$ for ${\cal E} = 140~ {\rm V/cm}$ and ${\cal B} = 40~ {\rm mG}$
is too small to give essential systematic error.


The work is supported by the Russian Science Foundation grant
No. 14-31-00022.

\begin{thebibliography}{14}
\expandafter\ifx\csname natexlab\endcsname\relax\def\natexlab#1{#1}\fi
\expandafter\ifx\csname bibnamefont\endcsname\relax
  \def\bibnamefont#1{#1}\fi
\expandafter\ifx\csname bibfnamefont\endcsname\relax
  \def\bibfnamefont#1{#1}\fi
\expandafter\ifx\csname citenamefont\endcsname\relax
  \def\citenamefont#1{#1}\fi
\expandafter\ifx\csname url\endcsname\relax
  \def\url#1{\texttt{#1}}\fi
\expandafter\ifx\csname urlprefix\endcsname\relax\def\urlprefix{URL }\fi
\providecommand{\bibinfo}[2]{#2}
\providecommand{\eprint}[2][]{\url{#2}}

\bibitem[{\citenamefont{Baron et~al.}(2014)\citenamefont{Baron, Campbell,
  Demille, Doyle, Gabrielse, Gurevich, Hess, Hutzler, Kirilov, Kozyryev
  et~al.}}]{Baron2013}
\bibinfo{author}{\bibfnamefont{J.}~\bibnamefont{Baron}},
  \bibinfo{author}{\bibfnamefont{W.~C.} \bibnamefont{Campbell}},
  \bibinfo{author}{\bibfnamefont{D.}~\bibnamefont{Demille}},
  \bibinfo{author}{\bibfnamefont{J.~M.} \bibnamefont{Doyle}},
  \bibinfo{author}{\bibfnamefont{G.}~\bibnamefont{Gabrielse}},
  \bibinfo{author}{\bibfnamefont{Y.~V.} \bibnamefont{Gurevich}},
  \bibinfo{author}{\bibfnamefont{P.~W.} \bibnamefont{Hess}},
  \bibinfo{author}{\bibfnamefont{N.~R.} \bibnamefont{Hutzler}},
  \bibinfo{author}{\bibfnamefont{E.}~\bibnamefont{Kirilov}},
  \bibinfo{author}{\bibfnamefont{I.}~\bibnamefont{Kozyryev}},
  \bibnamefont{et~al.}, \bibinfo{journal}{Science}
  \textbf{\bibinfo{volume}{343}}, \bibinfo{pages}{269} (\bibinfo{year}{2014}),
  ISSN \bibinfo{issn}{1095-9203}.

\bibitem[{\citenamefont{DeMille et~al.}(2001)\citenamefont{DeMille, Bay,
  Bickman, Kawall, Hunter, Krause, Maxwell, and Ulmer}}]{DeMille2001}
\bibinfo{author}{\bibfnamefont{D.}~\bibnamefont{DeMille}},
  \bibinfo{author}{\bibfnamefont{F.}~\bibnamefont{Bay}},
  \bibinfo{author}{\bibfnamefont{S.}~\bibnamefont{Bickman}},
  \bibinfo{author}{\bibfnamefont{D.}~\bibnamefont{Kawall}},
  \bibinfo{author}{\bibfnamefont{L.}~\bibnamefont{Hunter}},
  \bibinfo{author}{\bibfnamefont{D.}~\bibnamefont{Krause}},
  \bibinfo{author}{\bibfnamefont{S.}~\bibnamefont{Maxwell}}, \bibnamefont{and}
  \bibinfo{author}{\bibfnamefont{K.}~\bibnamefont{Ulmer}}, in
  \emph{\bibinfo{booktitle}{AIP Conference Proceedings}}
  (\bibinfo{publisher}{AIP}, \bibinfo{year}{2001}), vol. \bibinfo{volume}{596},
  pp. \bibinfo{pages}{72--83}, ISSN \bibinfo{issn}{0094243X},
  \urlprefix\url{http://link.aip.org/link/?APC/596/72/1&Agg=doi}.

\bibitem[{\citenamefont{Petrov et~al.}(2014)\citenamefont{Petrov, Skripnikov,
  Titov, Hutzler, Hess, O'Leary, Spaun, DeMille, Gabrielse, and
  Doyle}}]{Petrov:14}
\bibinfo{author}{\bibfnamefont{A.~N.} \bibnamefont{Petrov}},
  \bibinfo{author}{\bibfnamefont{L.~V.} \bibnamefont{Skripnikov}},
  \bibinfo{author}{\bibfnamefont{A.~V.} \bibnamefont{Titov}},
  \bibinfo{author}{\bibfnamefont{N.~R.} \bibnamefont{Hutzler}},
  \bibinfo{author}{\bibfnamefont{P.~W.} \bibnamefont{Hess}},
  \bibinfo{author}{\bibfnamefont{B.~R.} \bibnamefont{O'Leary}},
  \bibinfo{author}{\bibfnamefont{B.}~\bibnamefont{Spaun}},
  \bibinfo{author}{\bibfnamefont{D.}~\bibnamefont{DeMille}},
  \bibinfo{author}{\bibfnamefont{G.}~\bibnamefont{Gabrielse}},
  \bibnamefont{and} \bibinfo{author}{\bibfnamefont{J.~M.} \bibnamefont{Doyle}},
  \bibinfo{journal}{Phys. Rev. A} \textbf{\bibinfo{volume}{89}},
  \bibinfo{pages}{062505} (\bibinfo{year}{2014}).

\bibitem[{\citenamefont{Vutha and DeMille}(2009)}]{Vutha2009}
\bibinfo{author}{\bibfnamefont{A.}~\bibnamefont{Vutha}} \bibnamefont{and}
  \bibinfo{author}{\bibfnamefont{D.}~\bibnamefont{DeMille}},
  \bibinfo{journal}{arXiv}  (\bibinfo{year}{2009}), \eprint{0907.5116},
  \urlprefix\url{http://arxiv.org/abs/0907.5116}.

\bibitem[{\citenamefont{Petrov}(2015)}]{Petrov:15}
\bibinfo{author}{\bibfnamefont{A.~N.} \bibnamefont{Petrov}},
  \bibinfo{journal}{Phys.\ Rev.\ A} \textbf{\bibinfo{volume}{91}},
  \bibinfo{pages}{062509} (\bibinfo{year}{2015}).

\bibitem[{\citenamefont{Baron et~al.}(2017)\citenamefont{Baron, Campbell,
  Demille, Doyle, Gabrielse, Gurevich, Hess, Hutzler, Kirilov, Kozyryev
  et~al.}}]{ACME:17}
\bibinfo{author}{\bibfnamefont{J.}~\bibnamefont{Baron}},
  \bibinfo{author}{\bibfnamefont{W.~C.} \bibnamefont{Campbell}},
  \bibinfo{author}{\bibfnamefont{D.}~\bibnamefont{Demille}},
  \bibinfo{author}{\bibfnamefont{J.~M.} \bibnamefont{Doyle}},
  \bibinfo{author}{\bibfnamefont{G.}~\bibnamefont{Gabrielse}},
  \bibinfo{author}{\bibfnamefont{Y.~V.} \bibnamefont{Gurevich}},
  \bibinfo{author}{\bibfnamefont{P.~W.} \bibnamefont{Hess}},
  \bibinfo{author}{\bibfnamefont{N.~R.} \bibnamefont{Hutzler}},
  \bibinfo{author}{\bibfnamefont{E.}~\bibnamefont{Kirilov}},
  \bibinfo{author}{\bibfnamefont{I.}~\bibnamefont{Kozyryev}},
  \bibnamefont{et~al.} (\bibinfo{year}{2017}), \bibinfo{note}{arXiv:1612.09318
  [physics.atom-ph] (2017)}.

\bibitem[{\citenamefont{Skripnikov et~al.}(2013)\citenamefont{Skripnikov,
  Petrov, and Titov}}]{Skripnikov:13c}
\bibinfo{author}{\bibfnamefont{L.~V.} \bibnamefont{Skripnikov}},
  \bibinfo{author}{\bibfnamefont{A.~N.} \bibnamefont{Petrov}},
  \bibnamefont{and} \bibinfo{author}{\bibfnamefont{A.~V.} \bibnamefont{Titov}},
  \bibinfo{journal}{J.\ Chem.\ Phys.} \textbf{\bibinfo{volume}{139}},
  \bibinfo{eid}{221103} (\bibinfo{year}{2013}).

\bibitem[{\citenamefont{Skripnikov and Titov}(2015)}]{Skripnikov:15a}
\bibinfo{author}{\bibfnamefont{L.~V.} \bibnamefont{Skripnikov}}
  \bibnamefont{and} \bibinfo{author}{\bibfnamefont{A.~V.} \bibnamefont{Titov}},
  \bibinfo{journal}{The Journal of Chemical Physics}
  \textbf{\bibinfo{volume}{142}}, \bibinfo{eid}{024301} (\bibinfo{year}{2015}).

\bibitem[{\citenamefont{Skripnikov}(2016)}]{Skripnikov:16b}
\bibinfo{author}{\bibfnamefont{L.~V.} \bibnamefont{Skripnikov}},
  \bibinfo{journal}{J.\ Chem.\ Phys.} \textbf{\bibinfo{volume}{145}},
  \bibinfo{pages}{214301} (\bibinfo{year}{2016}).

\bibitem[{\citenamefont{Vutha et~al.}(2011)\citenamefont{Vutha, Spaun,
  Gurevich, Hutzler, Kirilov, Doyle, Gabrielse, and DeMille}}]{Vutha2011}
\bibinfo{author}{\bibfnamefont{A.~C.} \bibnamefont{Vutha}},
  \bibinfo{author}{\bibfnamefont{B.}~\bibnamefont{Spaun}},
  \bibinfo{author}{\bibfnamefont{Y.~V.} \bibnamefont{Gurevich}},
  \bibinfo{author}{\bibfnamefont{N.~R.} \bibnamefont{Hutzler}},
  \bibinfo{author}{\bibfnamefont{E.}~\bibnamefont{Kirilov}},
  \bibinfo{author}{\bibfnamefont{J.~M.} \bibnamefont{Doyle}},
  \bibinfo{author}{\bibfnamefont{G.}~\bibnamefont{Gabrielse}},
  \bibnamefont{and} \bibinfo{author}{\bibfnamefont{D.}~\bibnamefont{DeMille}},
  \bibinfo{journal}{Physical Review A} \textbf{\bibinfo{volume}{84}},
  \bibinfo{pages}{034502} (\bibinfo{year}{2011}), ISSN
  \bibinfo{issn}{1050-2947},
  \urlprefix\url{http://link.aps.org/doi/10.1103/PhysRevA.84.034502}.

\bibitem[{\citenamefont{Hess}(2014)}]{Hess2014Thesis}
\bibinfo{author}{\bibfnamefont{P.~W.} \bibnamefont{Hess}}, Ph.D. thesis,
  \bibinfo{school}{Harvard University}
  (\bibinfo{year}{2014}).

\bibitem[{\citenamefont{Landau and Lifshitz}(1977)}]{LL77}
\bibinfo{author}{\bibfnamefont{L.~D.} \bibnamefont{Landau}} \bibnamefont{and}
  \bibinfo{author}{\bibfnamefont{E.~M.} \bibnamefont{Lifshitz}},
  \emph{\bibinfo{title}{Quantum mechanics}} (\bibinfo{publisher}{Pergamon},
  \bibinfo{address}{Oxford}, \bibinfo{year}{1977}), \bibinfo{edition}{3rd} ed.

\bibitem[{\citenamefont{Edvinsson and Lagerqvist}(1984)}]{Edvinsson:84}
\bibinfo{author}{\bibfnamefont{G.}~\bibnamefont{Edvinsson}} \bibnamefont{and}
  \bibinfo{author}{\bibfnamefont{A.}~\bibnamefont{Lagerqvist}},
  \bibinfo{journal}{Physica Scripta} \textbf{\bibinfo{volume}{30}},
  \bibinfo{pages}{309} (\bibinfo{year}{1984}).

\bibitem[{\citenamefont{Petrov}(2011)}]{Petrov:11}
\bibinfo{author}{\bibfnamefont{A.~N.} \bibnamefont{Petrov}},
  \bibinfo{journal}{Phys.\ Rev.\ A} \textbf{\bibinfo{volume}{83}},
  \bibinfo{pages}{024502} (\bibinfo{year}{2011}).

\end{thebibliography}

\end{document}